\documentstyle[12pt]{article}
\newcommand{\be}{\begin{eqnarray}}
\newcommand{\ee}{\end{eqnarray}}

\catcode`\@=11
\def\lsim{\mathrel{\mathpalette\@versim<}}
\def\gsim{\mathrel{\mathpalette\@versim>}}
\def\@versim#1#2{\vcenter{\offinterlineskip
\ialign{$\m@th#1\hfil##\hfil$\crcr#2\crcr\sim\crcr } }}
\catcode`\@=12

\begin{document}

\begin{flushright}
KANAZAWA-03-01, IFUNAM-FT2003-01
 \\ January 2003
 \end{flushright}

 \begin{center}
{\Large\bf Renormalization Group Approach to \\
the SUSY Flavor Problem
\footnote{Talk given in the X Mexican School of
Particles and Fields, Playa del Carmen, M\' exico,
Oct. 30--Nov. 6, 2002}}
\end{center} 

\vspace{2cm}
\begin{center} {\sc Jisuke Kubo}
\end{center}

\begin{center}
{\em Instituto de F\' isica,  UNAM,
Apdo. Postal 20-364,
M\' exico 01000 D.F., M\' exico\\
and\\ Institute for Theoretical Physics,
Kanazawa  University, \\
Kanazawa 920-1192, Japan
}
\end{center}

\vspace{1cm}
\begin{center} {\bf \large Abstract}
 \end{center}
Renormalization group approach to
the SUSY flavor problem is elucidated.

\vspace{1cm}
\section{Introduction}
\subsection{Why  do we like supersymmetry?}
You certainly often have heard about what
supersymmetry (SUSY) is good for. Let me
mention few reasons for SUSY \cite{susy}.

The first one is the no-go theorem,
which was conjectured by Coleman and Mandula \cite{no-go}.
The theorem states:
No other  bosonic symmetries
of the S-matrix than the Poincar\' e invariance  and internal
symmetries are possible  in relativistic massive quantum field theories.
SUSY is the unique, nontrivial
extension of the Poincar\' e invariance.
This is why we would like that the super Poincar\'e invariance
is realized in nature. SUSY is a symmetry
between bosons and fermions, and predicts the existence
of superpartners with  different spins. Since however
no superpartners have been observed yet,
the super Poincar\' e invariance should be badly broken
at energies presently accessible to us, while
the Poincar\' e invariance
is exact so far. \footnote{A tinny violation was discussed by
G. Ramirez in this meeting, which might explain
the AGASA observation of the high-energy extragalactic
cosmic rays \cite{agasa}.}

The second reason for SUSY is that
SUSY can
stabilize the different mass scales.  In the Standard Model (SM),
for instance, there is a typical scale, which is the
mass of the Z boson. In the SM, there exists
an yet undiscovered particle, the Higgs particle, whose
mass is expected of the order of the weak scale, too.

It is also bounded from above due to the triviality
constraint, which was estimated to be about 0.6 TeV.
Then,  assume that the SM is embedded to a more
fundamental theory whose typical scale $\Lambda$ is  higher than
the weak scale. In the fundamental theory, there will be
heavy particles with mass of $O(\Lambda)$, and they will
contribute to the correction of the SM Higgs mass.
The correction  $\delta m^2_H$ will be
typically proportional to $(\alpha/\pi) \Lambda^2$,
where $\alpha$ is a generic coupling of the fundamental theory.
It would be ``unnatural'', if the correction $\delta m^2_H$ is
larger than the mass itself $m^2_H$ \cite{thooft,veltman}.
Since  $m_H$ is bounded from above,
we obtain the maximal natural scale of the SM, which is about few TeV,
where we assumed  $\alpha$ could be at most $O(1)$.

This means,  there will be
no natural extension of the SM
for energies larger than few TeV,
and consequently, there are no natural
grand unified theories (GUTs) \cite{thooft}.

SUSY can help a lot in this regard,
thanks to the non-renormalization theorem \cite{nonre}.
The theorem states:
There are only wave function renormalizations.
So, there is no extra renormalization for
mass parameters. In other words, there are no
quadratic divergences.
Therefore, the  correction $\delta m^2_H$ will be
proportional to $(\alpha/\pi) m_H^2 \ln \Lambda/m_H$
in a SUSY extension of the SM.
The requirement 
$\delta m^2_H < m_H^2$ is satisfied, even if
$\Lambda \sim M_{\rm PL}$ ( the Planck scale) .

The third reason for SUSY is experimental.
If the SM is unified to a SUSY GUT, it  predicts the unification of gauge
couplings of the SM. According to the report of Raby \cite{pdg},
the gauge unification in SUSY GUTS fits
to within $3 \sigma$ of the present precise  experimental data.
There are other  experimental indications:
Jens Erler and Wolgang Hollik reported in this meeting that
the  experimental values of
the W mass and  $g-2$ of the muon are   in favor of SUSY.
There are  other reasons for SUSY, but
they are often more subjective.

\subsection{How do we like SUSY to be broken?}
There are various 
mechanisms to break SUSY. \footnote{See for instance \cite{susy}.}
Although
we do not know which mechanism is
realized in nature, we would like SUSY to be broken
in such a way that it still can allow a natural
extension of the SM to a higher energy.
It has been studied how this requirement can be met,
and found that SUSY has to be only softly broken \cite{ssb}.
The soft-supersymmetry breaking (SSB) terms
are  those which do not change the infinity structure
of the parameters of the symmetric theory,
and do not produce quadratic divergences.
So they are additional
terms in the Lagrangian that do not change
the RG functions such as the $\beta$ and $\gamma$ functions
of the symmetric theory.
There exist four types of such terms \cite{ssb}:
$(m^2)_i^j \phi_j \phi^{* i}$
(soft scalar mass terms),
$B^{ij} \phi_i \phi_{j} $
($ B$-terms),
$M \lambda\lambda$ (gaugino mass terms),
$h^{ijk}\phi_i \phi_j\phi_k$
(trilinear scalar  couplings),
where $\phi_j$ and $\lambda$ denote the scalar component
in a chiral supermultiplet and the gaugino  in a gauge
supermultiplet, respectively.

The SSB parameters are massive parameters.
Although it depends on the model, we expect that
the SSB parameters are more than less in the same order, which
we call the SUSY breaking scale $M_{\rm SUSY}$.
To make compatible supersymmetry breaking
with the naturalness notion of  't Hooft and Veltman, we have to impose
the constraint on  $M_{\rm SUSY}$.
A simple calculation yields that $M_{\rm SUSY}$ should
be less than few TeV.

\subsection{What is the SUSY flavor problem?}
If we do not specify the SUSY breaking mechanism,
because we do not know which one is realized, and
rely only on renormalizability of the minimal extension
of the SM, the minimal supersymmetric SM (MSSM), it is possible to introduce
ciento cinco new parameters \cite{sutter}. \footnote{
The mass and mixing of neutrinos are not taken into account.}
So, all together we have ciento veinticuatro parameters!
 These 105 new parameters, of which 43 parameters are
CP violating phases, produce a lot of FCNC and CP violating
processes \cite{sutter}.

For instance, there exist diagrams due to the SSB term
$(m^2_{RL})^{e\mu}\tilde{e}_R \tilde{\mu}_L$ or
$(m^2_{LL})^{e\mu}(\tilde{e}_L)^* \tilde{\mu}_L$, which allow
the decay $\mu  \to e \gamma$ \cite{fcnc-mueg},
where $\tilde{e} (\tilde{\mu})$ denotes the scalarpartner
of the electron (muon).
The decay has not been observed yet.
A similar FCNC process is also possible in the hadronic sector,
which allows the decay $b \to s \gamma$ \cite{fcnc-bsg}.
Its branching ratio  ($=(3.3 \pm 0.4) \times 10^{-4}$)
has been measured at CLEO and BELLE \cite{cleo}, which
agrees well with the SM calculation
 ($=(3.29 \pm 0.33) \times 10^{-4}$) \cite{bsg}.
Renormalizability cannot explain why these SSB
parameters should be so small.
One of the most stringent constraint
on the CP-violation phases is the electric dipole moment
(EDM) of the neutron: $d_N/e \le 0(10^{-26}) cm $ \cite{pdg}.
The SSB term of the form
$(m^2_{LR})^{uu}\tilde{u}_L\tilde{u}_R$
can induce $d_N$, for instance, if
$(m^2_{LR})^{uu}$ is a complex parameter \cite{fcnc-edm}.
There exist a number of FCNC
and CP-violating processes,
 which give severe constraints on
the SSB parameters \cite{fcnc-mueg,fcnc-bsg,fcnc-edm,fcnc-k,fcnc}.

Fortunately, the  experimental constraints are not random:
They suggest that the soft scalar masses $(m^2)^i_j$
are diagonal in the space of generations
and the trilinear couplings
$h^{jik}$ are proportional to
the Yukawa couplings $Y^{ijk}$.
Although these  experimental constraints
hint a flavor symmetry in the SSB sector,
flavor symmetry is
 broken in nature and is not good enough
to overcome the SUSY flavor problem.
(Non-Abelian discrete symmetries such as $A_4$
could be used to  
 overcome the  problem\cite{babu}.)

\section{ What the RG approaches to the SUSY Flavor Problem
are based on.}
There exist various approaches to the SUSY Flavor Problem.
The  popular one is the hidden sector scenario.
In the so-called minimal supergravity model \cite{susy}
it is simply assumed    that
the SSB parameters have a universal form, say, at the GUT scale.
In this model, supersymmetry breaking occurs in a
sector that is hidden to the MSSM sector, and supersymmetry breaking
is mediated to the MSSM sector by gravity.
There exist other ideas of mediation:
gauge mediation \cite{gauge}, anomaly mediation \cite{anomaly} and gaugino
mediation \cite{gaugino}.

Another possibility is that
certain superparticles are so heavy
that FCNC and CP-violating processes may be  sufficiently
suppressed \cite{cohen}.
But this scenario could lead to problems, because, as we
 heard  from
Maria   Herrero and Maria Krawczyk

in this meeting, the heavy superparticles in the MSSM do not
necessarily decouple at low energies.
The approach that I am going to discuss below  is based on the
renormalization group (RG).
\subsection{Renormalizability and Infrared Attractiveness}
The main idea to solve the SUSY flavor problem
in the RG approach 
 is to use the infrared (IR) attractive
behavior of the SSB terms, which can  restrict the huge number of the
degrees of 
the freedom and then to increase the predictive power.
The basic idea goes back to Wilson \cite{wilson}, who
succeeded to relate IR and ultraviolet (UV) physics to
understand universality of critical phenomena.
Let me  illustrate in  a simple example
how renormalizability and IR attractiveness
are closely
related in the Wilsonian RG.
The example is
the scalar theory in three Euclidean dimension.
In the Wilsonian RG,
a set of infinite number of independent couplings
are allowed.
A RG flow is a trajectory in this infinite dimensional
space of couplings. However, all the trajectories
converge to a single trajectory,
called the renormalized trajectory (RT),  in the IR limit
for this example \cite{wilson}.

The RT trajectory intersects the critical surface at
the Wilson-Fisher fixed point.
The RT, on one hand,  defines the single coupling that
 survives in the IR limit, which
is identified with temperature in statistical physics.
The critical exponent expresses how fast a point
on the RT goes way from the fixed point.
To define a non-trivial relativistic field theory, that is,
nonperturbative renormalizability,
the couplings, on the other hand, have to lie on the RT. So,
all the couplings have to be functions of
this single independent coupling.
Of infinite number of couplings, there exists only one
independent coupling,
the same one in the IR as well as in the UV limit.

\subsection{Perturbative Renormalizability and Infrared Attractiveness}
The next question is:
Can one understand perturbative renormalizability
in terms of the Wilsonian RG?
This question was addressed by Polchinski \cite{polchinski},
who investigated 
RG trajectories in the scalar theory in four dimensions.
His observation, assuming that the mass parameter
can be neglected, is that there exists a  trajectory,
 which is attractive in the IR limit.
That is, whatever the initial values of the
unrenormalizable couplings  at
some UV scale are, they
become definite functions
of renormalizable couplings in the IR limit.
This is then interpreted as perturbative renormalizability.
Therefore, in the class of theories, in which
Polchinski's criterion \cite{polchinski} on perturbative renormalizability
can be applied, the UV cutoff
should be sufficiently large
so that physics in the infrared regime
has less dependence of
their initial values  in the UV regime.
\subsection{Pendleton-Ross fixed point and Zimmermann's
reduction of couplings}
There exists a further extension;
extension to relations among renormalizable  couplings.
The Pendleton-Ross IR fixed points \cite{pendleton}
are IR fixed points that one finds in various
theory models in the lower order in perturbation theory.
The IR fixed points have been used in
phenomenological approaches to various problems
in particle physics \cite{schrempp}. The approach emphasizes the
IR attractiveness of certain couplings.
The reduction of couplings of Zimmermann \cite{zimmermann} is
to actively reduce the number of
renormalizable couplings so that
it does not destroy perturbative renormalizability.
So, this approach emphasizes the UV
behavior of a theory. But as I have shown you in the Wilsonian RG, UV
and IR physics are closely related.
In fact, they give similar results.

The well known example is the SM.
If we neglect all the couplings except
the QCD gauge coupling $g_3$,
the top Yukawa coupling $Y_t$ and
the Higgs self coupling $\lambda_H$, one finds
at the one-loop level
that the ratios of the couplings
$(Y_t/g_3)^2$ and $(\lambda_H/g_3)^2$
approach to $2/9$ and $(\sqrt{689}-25)/18$ in the IR limit,
respectively \cite{pendleton}.
In fact these numbers are exactly the first coefficients in the
power series expansions
$(Y_t/g_3)^2=2/9+\sum_{n=1}\rho_n^t (g_3)^{2n}$ and
$(\lambda_H/g_3)^2=(\sqrt{689}-25)/18+\sum_{n=1}\rho_n^H (g_3)^{2n},$
which are the power series solutions to the so-called
reduction equations, and ensure perturbative renormalizability
to all orders in perturbation theory
in the reduced system of the SM \cite{kubo1}.
\subsection{Problems}
So, it is the obvious thing to try to apply these approaches to
softly broken SUSY 
theories \cite{ross2}. \footnote{There exist also works
based on the reduction of couplings \cite{kubo2}.}
The question is whether one can reduce
the huge number of the
independent SSB parameters of the MSSM
 in such a way that they do  not cause problems with the strong
experimental constraints.
The IR attractiveness results from
asymptotically free gauge interactions,
and fortunately, gauge interactions of the MSSM are
flavor independent.
It was found  \cite{ross2} in fact that they
are attracted to certain points in the IR limit,
and the points do not depend very much
on the flavor. But it turned out that the IR
attractiveness is not very strong.

To summarize, the IR attractiveness of the SSB
parameters in $4D$ dimensional softly broken Yang-Mills
(YM) theories is not strong enough that we
can satisfactory overcome the SUSY flavor problem.
The weakness of the IR attractiveness of the SSB
parameters in four dimension originates from
the fact that the scaling violation
 in four dimension are generally only logarithmic.
So, to achieve a stronger
 IR attractiveness of the SSB
parameters, we have to have a situation
in which the dimension is effectively or really more than four.
There are so far two ideas as far as I know.
The first one is based on a strongly coupled super
YM theory which has an IR fixed point.
The second one is to simply go to higher  dimensions.

\section{ Two viable solutions}
\subsection{Coupling to a superconformal gauge theory}
The first idea, proposed by Nelson and Strassler \cite{nelson},
 is to use a superconformal force
in a supersymmetric YM theory and rely on the
Seiberg conjecture \cite{seiberg}.
The conjecture states that a SUSY QCD-like YM with
has an IR fixed point in the so-called
conformal window \cite{banks,seiberg}.
As a result,   the anomalous dimensions for the matter
supermultiplets  become large of $O(1)$
near the IR fixed point.
Consequently, the parameters near the IR
fixed point run according to a  power-law, rather
than a logarithmic law.
 In a QCD-like theory we can not introduce the MSSM matter
multiplets. Fortunately, the conjecture
says that the QCD-like theory can be described by
another YM theory, which contains gauge singlet
mesons.  These mesons couples to the quark-like
matter  multiplets though the Yukawa coupling.
 The conjecture implies that  in the plane
of  the Yukawa and gauge couplings  there exists a nontrivial IR fixed
point.
Nelson and Strassler \cite{nelson}  identified the gauge singlet mesons
with the MSSM matter multiplets. In this way,
a large (in fact negative) anomalous dimensions are
transferred to the MSSM sector.
The hierarchy of the generations at low energies
could be achieved  in this way.

What happens now with the SSB parameters?
In \cite{karch}, it was  shown that the SSB parameters
approach zero in the IR limit
both in the original and dual theories.
What about the SSB parameters
of the MSSM? Since the matter and Higgs supermultiplets
directly couple to the (dual) superconformal
sector, the corresponding
SSB parameters, i.e., $m^2$'s and $h$'s,
will approach zero, while the gaugino masses
remain as they are.
Then the scenario to solve the SUSY flavor problem is:
There are two scales $  M_>$ and $  M_<$  in addition to $ M_{\rm SUSY}$.
Above $M_>$
(which may be between $10^{14}$ to $10^{19}$  GeV)
 the strong sector is perturbative.
Between$ M_>$ and $M_<$, the theory is in the conformal
regime, and the SSB parameters should obey the power
law running, and become  almost  zero at $M_<$.
Below $M_<$ (which may be between $10^{10}$ to $10^{16}$  GeV),
 the superconformal
sector escapes by some mechanism, that is, the MSSM sector
decouples from  the superconformal sector, the MSSM interactions,
especially the gauge interactions,
generate   $m^2$'s and $h$'s of the
MSSM  at $ M_{\rm SUSY}$.
Remember, the MSSM gauge interactions are flavor
independent. Therefore, they can
 generate flavor independent SSB parameters at low energies \cite{nelson},
although the flavor independence
is not perfect and it could be a
problem \cite{kobayashi0}.
There  are further extensions
and 
applications of the idea of Nelson and Strassler \cite{kobayashi,kubo4}.
\subsection{Going to higher dimensions}
The next idea is to really  go to higher dimensions
\cite{antoniadis,arkani}.
In higher dimensions,
due to  the power-law running of
couplings \cite{veneziano,dienes,kobayashi1,ejiri},
stronger infrared attractiveness \cite{abel} of
the SSB 
parameters is expected \cite{nunami,espinoza,kubo3,choi}.
In \cite{kubo3} we considered the simplest case in which only
the gauge supermultiplet
propagates in the $(4+\delta)$-dimensional bulk and the supermultiplets
containing the matter and Higgs fields are localized
at our 3-brane \cite{antoniadis,arkani,dienes}.
The gaugino mass $M$,
which is assumed to be generated at the fundamental scale $M_{\rm PL}$
by some SUSY breaking mechanism,
receives
a correction proportional to
$(M_{\rm PL}/M_{\rm GUT})^\delta$
at the grand unification scale $M_{\rm GUT}$,
and more importantly it  can induce flavor-blind corrections to other
SSB parameters.
We found \cite{kubo3} that the soft scalar masses
$(m^2)^i_j$
and the soft-trilinear couplings $h^{ijk}$
become  so aligned at $M_{\rm GUT}$
that FCNC processes and
dangerous CP-violating phases are sufficiently
suppressed (see also \cite{choi}):
\begin{eqnarray}
h^{ijk} / M  Y^{ijk} = -\eta_Y^{ijk}+O(10^{-6})~,~
(m^2)^{i}_{j}/ |M|^2 = C(i)/C(G) \delta_j^i+O(10^{-3}),
\nonumber
\end{eqnarray}
at $M_{\rm GUT}$,
if we take into account only
power law corrections,
where  $Y^{ijk}$
are Yukawa couplings, $C$'s and $\eta$'s are
grouptheoretic constants, and
we assumed that 
$ M_{\rm PL}/M_{\rm GUT} =10^2$ and $\delta=2$.

As a concrete example, we \cite{kubo3} considered
the minimal
supersymmetric SU(5) GUT model in six dimensions including logarithmic
corrections, and found that
all the A-parameters $h$'s,
B-parameter $B$ and  soft-scalar masses $m^2$'s
for the MSSM
are  fixed  once
 the unified gaugino mass $M$ is given.
For instance, $\tan \beta \simeq 19.5$ for
$M=0.5$ TeV.

Nevertheless, 
the  stringent constraints coming from
the $K_S-K_L$ mass difference $\Delta m_K$,
$\epsilon'/\epsilon$ in the
$K^0-\bar{K^0}$ mixing, the decay
$\mu\to e\gamma$, and
the electric dipole moments (EDMs) of the neutron and the
electron \cite{fcnc} are satisfied in this model.

The suppression mechanism of the FCNC and CP-phases presented
above does not work in four dimensions. Therefore, the smallness of FCNC
as well as of EDM is a possible hint of the  existence of extra dimensions.

\begin{center} Acknowledgments
    \end{center}
 I would like to thank the organizers
of  the X Mexican School of Particles and Fields
for offering me a chance to
give a talk in such a beautiful conference location
in a very special occasion which is devoted
to two important 
Latin American physicists, Augusto Garc\' ia and
Arnulfo Zepeda.

\end{document}